\documentclass[11pt,a4paper]{article}

\usepackage[utf8]{inputenc}
\usepackage[T1]{fontenc}
\usepackage{mathptmx}
\usepackage{helvet}
\usepackage{courier}
\usepackage[margin=1in]{geometry}
\usepackage{setspace}
\usepackage{graphicx}
\usepackage{booktabs}
\usepackage{tabularx}
\usepackage{array}
\usepackage{multirow}
\usepackage{enumitem}
\usepackage{hyperref}
\usepackage{xcolor}
\usepackage{titlesec}
\usepackage[numbers,square]{natbib}
\usepackage{authblk}
\hypersetup{colorlinks=true,linkcolor=black,citecolor=black,urlcolor=blue}

\newcolumntype{L}[1]{>{\raggedright\arraybackslash}p{#1}}

\graphicspath{{figs/}}

\titleformat{\section}{\Large\bfseries}{\thesection}{1em}{}
\titleformat{\subsection}{\large\bfseries}{\thesubsection}{1em}{}
\titleformat{\subsubsection}{\normalsize\bfseries\itshape}{\thesubsubsection}{1em}{}

\title{\vspace{-1.5em}\Large\bfseries Towards an Ethical AI Curriculum:\\
A Pan-African, Culturally Contextualized Framework\\ for Primary and Secondary Education}

\author[1]{Abidemi Kuburat Adedeji\thanks{Corresponding author: \href{mailto:adedejibdm@gmail.com}{adedejibdm@gmail.com}}}
\author[2]{Franklin Tchakount\'{e}}
\author[3]{Sulaiman Oluwasegun Yusuff}

\affil[1]{Department of Computer Science, Abraham Adesanya Polytechnic, Ijebu-Igbo, Ogun, Nigeria}
\affil[2]{Department of Mathematics and Computer Science, University of Ngaound\'{e}r\'{e}, Dang, Adamaoua 454, Cameroon}
\affil[3]{Department of Computer Science, Ball State University, Muncie, IN 47306, USA}

\date{April 29, 2026}

\begin{document}

\maketitle

\
oindent\\textit{Preprint. Submitted to AI and Ethics (Springer Nature).}\\vspace{0.5em}

\begin{abstract}
Artificial intelligence (AI) is now embedded in educational, civic, and economic systems worldwide. For African primary and secondary education, this creates a double imperative: to prepare a young population---over sixty per cent of Africans are under twenty-five---for AI-mediated labour markets, and to do so without uncritically importing curricula designed for other linguistic, cultural, and socio-political contexts. The African Union's Continental AI Strategy, adopted in July 2024, and the 2025 Africa Declaration on AI have elevated these questions to the continental agenda. This paper proposes a Pan-African, culturally contextualised, and ethically grounded framework for integrating AI education into African primary and secondary schools. Methodologically, the paper is a structured conceptual synthesis of continental and national policy documents (African Union, UNESCO, and national strategies including Kenya's National AI Strategy 2025--2030), peer-reviewed scholarship on AI ethics, AI literacy, decolonial pedagogy, and Ubuntu-grounded AI governance, and institutional perspectives from UNESCO's AI Readiness Assessments and Smart Africa. We do not report empirical classroom data; rather, we contribute (i)~a framework comprising six guiding principles, four curriculum domains, five ethical competencies, and an age-banded progression from lower primary to upper secondary; (ii)~a comparative analysis of continental and national policy contexts; (iii)~an explicit mapping between global AI-ethics principles and Ubuntu-informed relational ethics; (iv)~a planned empirical validation programme combining a Delphi study with African AI-ethics scholars, teacher surveys across anglophone, francophone, lusophone, and arabophone contexts, and multi-country classroom piloting; and (v)~targeted recommendations for policymakers, educators, civil society, and international partners. We argue that an ethical AI curriculum can serve as a transformative tool for equity, innovation, and social justice, and we outline a research agenda to embed ethics, resilience, and critical thinking at the core of Africa's digital future.

\vspace{0.6em}\noindent\textbf{Keywords:} AI literacy; AI ethics; African education; curriculum design; decolonial pedagogy; Ubuntu; culturally responsive teaching; digital inclusion; African Union Continental AI Strategy; UNESCO Recommendation on the Ethics of AI; teacher professional development; youth empowerment.
\end{abstract}

\section{Introduction}
Artificial intelligence is rapidly transforming education systems worldwide, offering new possibilities for personalised learning, administrative efficiency, and student engagement~\citep{wang2023aiedu,unesco2021aiguidance}. Across Africa, where more than sixty per cent of the population is under twenty-five, governments and stakeholders increasingly recognise AI's potential to bridge educational gaps and prepare learners for participation in an AI-driven global economy~\citep{au2024contai,emeje2023reframing}. Two recent developments have reframed the conversation. The African Union's Continental AI Strategy, endorsed at the 45th Executive Council in Accra in July 2024, commits member states to a people-centred, development-oriented approach to AI and foregrounds education as a pillar of inclusive digital futures~\citep{au2024contai}. In April 2025, forty-nine African countries signed the Africa Declaration on AI alongside a proposed sixty-billion-dollar Africa AI Fund~\citep{africa2025declaration}. These commitments create both opportunity and risk: opportunity to build curricula aligned with African realities, and risk of uncritical adoption of imported curricula that reproduce what \citet{birhane2021algorithmic} describes as algorithmic colonialism.

Current AI curricula are predominantly developed in the Global North and frequently overlook Africa's linguistic, cultural, and socio-political contexts~\citep{birhane2021algorithmic,kizilcec2022fairness}. Uncritical adoption risks reinforcing digital colonialism, cultural dissonance, and the erasure of indigenous knowledge systems~\citep{birhane2021algorithmic,adams2023postcolonial}. It also tends to foreground technical content at the expense of ethical, civic, and critical dimensions. Failure to address language diversity, gender disparities, and localised ethical concerns can deepen inequity and reduce the efficacy of AI literacy initiatives~\citep{chaudhary2021gender,williams2020equity}. A culturally contextualised, ethically grounded AI curriculum framework is therefore not merely desirable; it is imperative.

\subsection{Urgency of Preparing African Learners for AI-Mediated Economies}
Africa's youthful demographic is projected to reach approximately 1.7~billion by 2030~\citep{au2024contai,worldbank2023ai}. AI literacy---encompassing both technical and ethical understanding---is foundational for workforce readiness across sectors including agriculture, healthcare, finance, and governance~\citep{emeje2023reframing,worldbank2023ai}. Without deliberately crafted, ethically informed AI education, many African learners risk marginalisation in rapidly evolving labour markets and technological landscapes~\citep{kizilcec2022fairness,williams2020equity}.

\subsection{Risks of Importing AI Curricula Without Contextual Adaptation}
Most existing AI education frameworks have been developed in high-income contexts whose socio-economic, cultural, and linguistic assumptions differ markedly from those across Africa~\citep{birhane2021algorithmic,kizilcec2022fairness}. Uncritical adoption risks reinforcing digital colonialism, cultural dissonance, and erasure of indigenous knowledge systems~\citep{adams2023postcolonial,mhlambi2020ubuntu}. It also tends to centre Western-specific ethical vocabularies, underplaying African ethical traditions such as Ubuntu~\citep{mhlambi2020ubuntu,ogoh2025ubuntu}.

\subsection{Purpose of the Study}
This study proposes a Pan-African, culturally contextualised, and ethically grounded AI curriculum framework tailored to the realities of African primary and secondary education systems. The framework empowers learners as informed and ethical AI users and co-creators by embedding indigenous knowledge, ethical reasoning, and inclusive pedagogy. Beyond the framework itself, the paper identifies structural enablers and barriers, offers stakeholder-targeted recommendations, and outlines an empirical validation programme.

\subsection{Research Questions}
Guided by the challenges and opportunities identified, this study addresses four core research questions:
\begin{itemize}[leftmargin=*,itemsep=2pt]
\item \textbf{RQ1.} What constitutes an ethical AI curriculum for African primary and secondary learners, as distinct from generic AI literacy?
\item \textbf{RQ2.} Which values, competencies, and pedagogical strategies are essential for meaningful, culturally relevant AI education in Africa?
\item \textbf{RQ3.} How can such a curriculum be adapted and implemented across Africa's diverse linguistic, cultural, and infrastructural contexts?
\item \textbf{RQ4.} How might the proposed framework be empirically validated through Delphi, survey, and classroom-based pilot studies?
\end{itemize}

\section{Background and Related Work}
Artificial intelligence is increasingly shaping societies across economic, cultural, and educational domains. Globally, countries are embedding AI literacy into formal education, with substantial variation in scope and ethical emphasis~\citep{wang2023aiedu,unesco2021aiguidance}. Critiques note that high-income models frequently marginalise ethical reasoning, cultural inclusion, and civic engagement~\citep{jobin2019global,birhane2021algorithmic,adams2023postcolonial}.

\subsection{The Continental Policy Context}
The African Union's Continental AI Strategy (2024) identifies five focus areas and fifteen action areas, foregrounding ``people-centric, development-oriented and inclusive'' AI~\citep{au2024contai}. Its implementation timeline extends to 2030, with Phase~1 (2025--2026) establishing governance structures and national strategies. The earlier Digital Transformation Strategy for Africa (2020--2030) positioned education as a critical enabler of inclusive digital futures but offered limited operational guidance for AI literacy~\citep{smartafrica2022ai}. UNESCO's Recommendation on the Ethics of Artificial Intelligence, adopted in 2021, remains the principal international standard and includes an ``Ubuntu paragraph'' recognising the relational dimensions of personhood~\citep{unesco2021recom,unesco2023genai}. More recently, UNESCO has piloted AI Readiness Assessment Methodology (RAM) studies in Southern Africa and supported AU-aligned work on AI in education~\citep{unesco2024ram,unesco2024iicba}.

National initiatives vary in scope. Kenya's National AI Strategy 2025--2030, published in March 2025, commits to embedding AI literacy across education and to piloting Digital Innovation Hubs (DigiKen) as curriculum anchors~\citep{kenya2025ai,ouma2024kenya,team4tech2025kenya}. South Africa has invested in coding and robotics in selected grades. Ghana and Rwanda have announced AI and digital-skills strategies linked to Smart Africa~\citep{smartafrica2022ai}. Nigeria's National Digital Economy Policy and Strategy emphasises population-wide digital literacy~\citep{nitda2020ndeps}, while Cameroon integrates ICT into secondary programmes~\citep{cameroon2023ict}. Each initiative faces distinctive infrastructural and policy challenges~\citep{team4tech2025kenya,wilson2024regulating}.

\subsection{AI Ethics Frameworks Relevant to Education}
The AI ethics literature converges on a core set of principles---fairness, transparency, accountability, privacy, and inclusivity---though interpretations and emphases differ across contexts~\citep{jobin2019global,unesco2021recom}. Recent African scholarship stresses that these principles must be interpreted through relational and decolonial lenses. Ubuntu---the African philosophy of interconnectedness captured by ``I am because we are''---has been advanced as both a complement and a corrective to Western-centric AI ethics~\citep{mhlambi2020ubuntu,ogoh2025ubuntu,jecker2024ubuntu}. \citet{mhlambi2020ubuntu} reframes rationality as relationality; \citet{metz2007african} provides a moral-philosophical grounding; and recent work locates Ubuntu within AI governance~\citep{ogoh2025ubuntu,jecker2024ubuntu}.

\subsection{AI Literacy and Teacher Capacity in Africa}
Empirical studies of teacher preparedness in African contexts consistently identify teacher capacity as the single strongest determinant of successful integration~\citep{ouma2024kenya,williams2024teacher}. In Kenya, a 2024 study found that AI is not yet part of the teacher-training curriculum, with limited professional development opportunities for pre-service and in-service teachers~\citep{kenya2025ai,ouma2024kenya}. African AI research communities---Masakhane, Deep Learning Indaba, Data Science Africa, and Zindi---have built vibrant networks that can feed culturally relevant content and teacher capacity-building~\citep{nekoto2020masakhane,africanlp2024}.

\subsection{Gap Addressed and Theoretical Novelty}
Existing frameworks either (a)~address AI literacy without ethical or cultural grounding, (b)~articulate AI ethics in abstract or Global-North-centric terms, or (c)~describe African digital-skills policies without translating them into classroom-ready curriculum design. \citet{selwyn2025unesco} warn against techno-solutionism in UNESCO's guidance on AI in education. Relatedly, the African Observatory on Responsible AI and the African Union High-Level Panel on AI have called for pedagogical work that operationalises continental ethics commitments at the classroom level. This paper addresses the gap by proposing a Pan-African curriculum framework centred on ethics, Ubuntu-informed relational reasoning, indigenous knowledge, and inclusion, with explicit age-banded progression, continental policy grounding, and an empirical validation programme.

The theoretical novelty of the framework is threefold. First, it explicitly operationalises the complementarity between global AI-ethics principles and African relational ethics, rather than treating Ubuntu as a cultural decoration layered onto Global-North content. Second, it weaves ethical, technical, and pedagogical progression together across four age bands, whereas most existing models treat ethics as an afterthought. Third, it couples framework articulation with a concrete, ethics-reviewed empirical validation programme (Section~\ref{sec:empirical}), closing the gap between policy rhetoric and classroom evidence that has characterised much Global-South AI-education work.

\section{Methodology}
This study is a structured conceptual synthesis. It integrates three complementary evidence streams---policy documents, peer-reviewed literature, and institutional guidance---into a coherent curriculum framework.

\subsection{Design Approach}
We adopt a thematic synthesis approach~\citep{thomas2008thematic}, widely used in educational policy research. The approach proceeds in four stages: (a)~scoping the relevant policy and literature corpus; (b)~extracting themes related to ethical principles, competency domains, and pedagogical strategies; (c)~clustering themes into a coherent framework; and (d)~cross-checking the framework against African educational realities reported in the literature.

\subsection{Sources Analysed}
\begin{enumerate}[leftmargin=*,itemsep=2pt]
\item \emph{Continental and international policy.} The African Union Continental AI Strategy (2024)~\citep{au2024contai}; the Africa Declaration on AI (2025)~\citep{africa2025declaration}; the AU Digital Transformation Strategy for Africa (2020--2030)~\citep{smartafrica2022ai}; UNESCO's Recommendation on the Ethics of Artificial Intelligence (2021)~\citep{unesco2021recom}; UNESCO guidance on generative AI in education and research~\citep{unesco2023genai}; and UNESCO's AI Readiness Assessments in Southern Africa~\citep{unesco2024ram,unesco2024iicba}.
\item \emph{National strategies and education frameworks.} Illustrative documents from Kenya, South Africa, Ghana, Rwanda, Nigeria, and Cameroon, reviewed for treatment of ethics, curriculum, teacher capacity, and language.
\item \emph{Peer-reviewed scholarship and African research-community outputs.} AI ethics~\citep{birhane2021algorithmic,mhlambi2020ubuntu,ogoh2025ubuntu,jobin2019global,metz2007african}; AI literacy and education~\citep{wang2023aiedu,unesco2021aiguidance,emeje2023reframing,chaudhary2021gender,williams2020equity}; decolonial and culturally responsive pedagogy~\citep{birhane2021algorithmic,adams2023postcolonial,freire2000pedagogy,vygotsky1978mind,gay2018culturally}; Ubuntu philosophy~\citep{mhlambi2020ubuntu,ogoh2025ubuntu,jecker2024ubuntu,metz2007african}; and African AI research communities~\citep{nekoto2020masakhane,africanlp2024}.
\end{enumerate}

\subsection{Thematic Synthesis}
Codes were generated inductively from the sources and organised into three analytical categories: (a)~values, (b)~competencies, and (c)~pedagogy. A fourth category---(d)~implementation conditions---emerged during coding and was preserved because it proved indispensable for translating framework into practice.

\subsection{Limitations of the Methodology}
As a conceptual synthesis, this study does not include primary empirical data. The framework is therefore a hypothesis for practice rather than an evaluated intervention. An empirical validation programme is proposed in Section~\ref{sec:empirical}. The Pan-African scope is necessarily aspirational. The source base also skews towards English-language scholarship; francophone, lusophone, and arabophone sources are under-represented and will be strengthened in the empirical phase.

\section{Theoretical Framework}
\begin{figure}[ht]
\centering
\includegraphics[width=0.85\textwidth]{eth_fig1_framework.png}
\caption{The Pan-African Ethical AI Curriculum Framework in concentric layers: cultural grounding (Ubuntu, indigenous knowledge, African languages), pedagogical approach (critical, constructivist, culturally responsive, transformative), and a core of five AI-ethics principles, four curriculum domains, and five ethical competencies.}
\label{fig:framework}
\end{figure}

\subsection{Ethical AI Principles Relevant to Education}
Five principles anchor the framework's ethical core: fairness, transparency, accountability, privacy, and inclusivity~\citep{unesco2021aiguidance,jobin2019global,unesco2021recom}. In educational settings these principles take specific pedagogical form: fairness demands that AI-related learning activities and assessments avoid bias and promote equitable access; transparency requires that learners can interrogate how AI systems reach conclusions at an age-appropriate level; accountability ensures institutional mechanisms of recourse for algorithmic harms; privacy protects sensitive learner data; inclusivity stresses culturally responsive and accessible AI learning~\citep{emeje2023reframing,gay2018culturally}. Table~\ref{tab:ubuntu} maps each principle to Ubuntu-informed relational ethics.

\begin{table}[ht]
\centering
\caption{Mapping of five AI-ethics principles to Ubuntu-informed relational-ethics extensions for African classrooms.}
\label{tab:ubuntu}
\small
\begin{tabularx}{\textwidth}{@{}lXX@{}}
\toprule
\textbf{AI-ethics principle} & \textbf{Standard interpretation} & \textbf{Ubuntu-informed extension} \\ \midrule
Fairness & Avoidance of statistical bias; equality of opportunity & Relational fairness: effects on community members; inter-generational equity; restorative orientation \\
Transparency & Explainability of model outputs & Accountability to the community for how decisions are made; explanation in learners' own languages \\
Accountability & Legal or institutional recourse & Shared responsibility across learners, teachers, families, and community leaders; mutual obligation \\
Privacy & Data protection and consent & Collective and familial dimensions of data; stewardship of community knowledge \\
Inclusivity & Access and representation & Belonging within a greater whole; participation across gender, ability, language, rural--urban divides \\ \bottomrule
\end{tabularx}
\end{table}

\subsection{Pedagogical Foundations}
Four pedagogical traditions align well with ethical AI education. Critical pedagogy and reflexivity foreground the social and political dimensions of technology~\citep{freire2000pedagogy}. Constructivist theory emphasises active, inquiry-based engagement in which learners explore the societal impacts of AI rather than passively consuming technical content~\citep{vygotsky1978mind}. Culturally responsive pedagogy centres learners' cultural backgrounds and experiences, vital for contextualising AI ethics in African classrooms~\citep{gay2018culturally}. Transformative learning theory fosters the ethical reasoning and digital agency learners need to navigate AI responsibly~\citep{mezirow1997transformative}.

\subsection{Culturally Responsive and Inclusive Education Models}
Culturally responsive education integrates indigenous knowledge, local languages, and community values, and challenges dominant Western paradigms~\citep{emeje2023reframing,adams2023postcolonial}. Ubuntu---emphasising relationality, mutual respect, and interdependence---provides a specifically African ethical grounding that complements the five AI-ethics principles~\citep{mhlambi2020ubuntu,ogoh2025ubuntu,jecker2024ubuntu,metz2007african}. Inclusive education principles further ensure curricula address gender equity, disability access, language diversity, and socio-economic barriers~\citep{unesco2020inclusion}.

\subsection{An Integrated Foundation}
Global AI-ethics principles are necessary but must be interpreted through critical, constructivist, and culturally responsive pedagogies and grounded in African ethical thought. This integration distinguishes the framework proposed here from generic AI-literacy curricula.

\section{A Pan-African Ethical AI Curriculum Framework}

\subsection{Guiding Principles}
The framework rests on six guiding principles that bridge ethical AI, African values, and inclusive education:
\begin{enumerate}[leftmargin=*,itemsep=2pt]
\item \emph{Local agency.} Learners, teachers, and communities co-design AI education rather than passively receive imported content.
\item \emph{Ethical grounding.} AI ethics is foundational rather than supplementary.
\item \emph{Multilingual inclusivity.} Content is delivered in national and widely spoken African languages as well as English, French, Portuguese, and Arabic.
\item \emph{Cultural responsiveness.} Local knowledge systems anchor technical concepts.
\item \emph{Equity and inclusion.} Explicit attention to gender, disability, rural--urban divides, and socio-economic disparities shapes design and delivery.
\item \emph{Critical digital citizenship.} Learners engage with AI not only as users or builders but as citizens capable of scrutinising its societal effects.
\end{enumerate}

\subsection{Curriculum Domains}
The framework organises content into four interrelated domains:
\begin{itemize}[leftmargin=*,itemsep=2pt]
\item \emph{Conceptual foundations of AI.}
\item \emph{Data literacy and critical thinking.}
\item \emph{Ethical and societal dimensions.}
\item \emph{Creative and civic application.}
\end{itemize}

\subsection{Age-Banded Progression}
\begin{figure}[ht]
\centering
\includegraphics[width=0.95\textwidth]{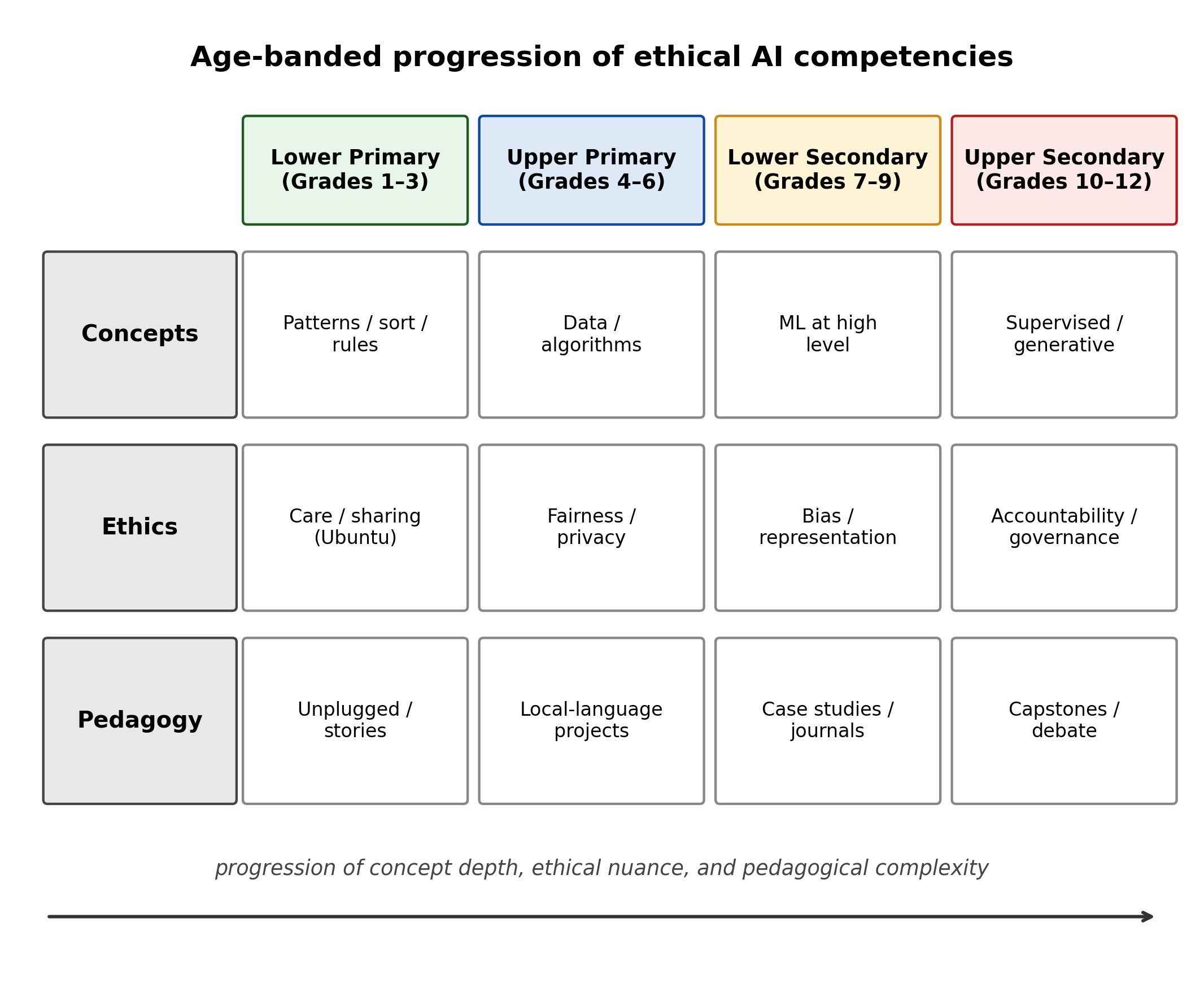}
\caption{Age-banded progression of concepts, ethics, and pedagogy. Concept depth, ethical nuance, and pedagogical complexity deepen together rather than sequentially.}
\label{fig:progression}
\end{figure}

\begin{table}[ht]
\centering
\caption{Age-banded progression.}
\label{tab:bands}
\small
\begin{tabularx}{\textwidth}{@{}lXXX@{}}
\toprule
\textbf{Band} & \textbf{Core Concepts} & \textbf{Ethical Focus} & \textbf{Pedagogy \& Activities} \\ \midrule
Lower Primary (1--3) & What is AI? Patterns, sorting, rules. Humans vs.\ machines. & Honesty, sharing, care for others and environment (Ubuntu). & Unplugged activities; storytelling with local folktales; simple pattern games. \\
Upper Primary (4--6) & Data, instructions, simple algorithms. Examples from farming, weather, language. & Fairness in rules; whose data is it? Privacy at home and at school. & Group projects with community data; role-plays; discussions in local languages. \\
Lower Secondary (7--9) & Machine learning at a high level; chatbots; image and speech tools. & Bias, stereotypes, representation in data. Who benefits? Who is harmed? & Case studies from African contexts; critical analysis of AI products; reflective journals. \\
Upper Secondary (10--12) & Supervised learning, generative models, evaluation, limitations of AI. & Accountability, governance, algorithmic colonialism, environmental and labour costs. & Capstone projects co-designed with community partners; debate and policy analysis. \\ \bottomrule
\end{tabularx}
\end{table}

\subsection{Ethical Competency Areas}
Across bands, the framework develops five ethical competencies (``recognise, interrogate, evaluate, reason, act''). Table~\ref{tab:comp} sketches the progression of expected learner outcomes.

\begin{table}[ht]
\centering
\caption{Progression of expected learner outcomes across five ethical competencies.}
\label{tab:comp}
\scriptsize
\begin{tabularx}{\textwidth}{@{}lXXXX@{}}
\toprule
\textbf{Competency} & \textbf{Lower Primary} & \textbf{Upper Primary} & \textbf{Lower Secondary} & \textbf{Upper Secondary} \\ \midrule
Recognise AI in daily life & Identify everyday ``smart'' tools & Describe what these tools do & Explain inputs, outputs, and limits & Analyse system design choices \\
Interrogate data & Ask where information comes from & Describe bias in examples & Evaluate datasets for inclusion & Audit data for representation gaps \\
Evaluate outputs & Spot obvious errors & Compare AI output to real-world knowledge & Use accuracy, fairness criteria & Propose evaluation protocols \\
Reason ethically & Apply Ubuntu values in class scenarios & Apply fairness/privacy to AI cases & Reason about harms and trade-offs & Construct full ethical arguments \\
Act responsibly & Choose to share / not share data & Use AI tools considerately & Design respectful AI experiments & Advocate for change in communities \\ \bottomrule
\end{tabularx}
\end{table}

\subsection{Pedagogical Strategies}
Teaching strategies consistent with this framework include inquiry-based learning, problem-based projects grounded in local realities, Socratic discussion around case studies, structured debates, role-play, and capstone projects co-designed with community partners. Unplugged activities are particularly important in infrastructure-constrained schools~\citep{team4tech2025kenya,williams2020equity}.

\subsection{Language and Cultural Localisation}
Materials should be designed for delivery in at least three registers: the official language(s) of instruction; widely spoken African lingua francas; and vernacular languages. African NLP resources from Masakhane and partners~\citep{nekoto2020masakhane,africanlp2024} offer a practical starting point for translation and content adaptation.

\subsection{Resource Profile by Band}
\begin{table}[ht]
\centering
\caption{Resource, bandwidth, and teacher-PD profile by age band.}
\label{tab:resource}
\small
\begin{tabularx}{\textwidth}{@{}lXXX@{}}
\toprule
\textbf{Band} & \textbf{Device dependence} & \textbf{Bandwidth needs} & \textbf{Teacher PD} \\ \midrule
Lower Primary (1--3) & None required; optional tablet demo & None for core activities & 2--3 days foundational \\
Upper Primary (4--6) & Shared device / rotation & Low (offline-first) & 3--5 days + community of practice \\
Lower Secondary (7--9) & Shared device; periodic online & Moderate (intermittent) & 5--7 days + mentorship \\
Upper Secondary (10--12) & Regular device access; periodic cloud & Moderate--high during projects & Ongoing PD + capstone supervision \\ \bottomrule
\end{tabularx}
\end{table}

\subsection{Assessment and Evaluation}
Assessment should be multidimensional and triangulated: knowledge checks for conceptual content; project rubrics for data literacy and creative application; reflective portfolios and oral assessments for ethical reasoning and digital citizenship; and teacher and peer observation for collaboration and inclusivity.

\section{Implementation Considerations and Enablers}

\subsection{Teacher Capacity}
Teacher professional development is the single strongest determinant of curriculum impact in African contexts~\citep{kenya2025ai,ouma2024kenya,williams2024teacher}. Effective rollout requires structured in-service training on AI concepts, AI ethics, and the pedagogical strategies described above; communities of practice that support peer learning; and recognition mechanisms that value ethical AI teaching alongside technical instruction.

\subsection{Infrastructure and Access}
Infrastructural disparities remain a defining constraint~\citep{williams2020equity}. The Kenyan 2024 baseline reports about 35\% of surveyed schools had working internet and that data costs remain high~\citep{kenya2025ai,team4tech2025kenya}. The framework is intentionally designed to be low-bandwidth and device-optional at lower bands.

\subsection{Policy and Governance}
National ministries should co-develop AI curriculum standards with the African Union Commission and UNESCO~\citep{au2024contai,unesco2021recom,unesco2024ram}. Governance mechanisms should include explicit data-protection provisions for learners, guidelines on the classroom use of generative AI~\citep{unesco2023genai}, and procurement standards that protect schools from extractive data practices.

\subsection{Civil Society and Youth-Led Engagement}
Civil society organisations, educators' unions, parent and community associations, and youth-led digital-rights groups bring critical perspective and legitimacy to curriculum design~\citep{birhane2021algorithmic,adams2023postcolonial}.

\subsection{Partnerships}
International partners can play a constructive role when partnerships are rooted in African leadership, transparent governance, and equitable arrangements. African AI research communities~\citep{nekoto2020masakhane,africanlp2024} are indispensable partners.

\subsection{Stakeholder Ecosystem}
\begin{figure}[ht]
\centering
\includegraphics[width=0.70\textwidth]{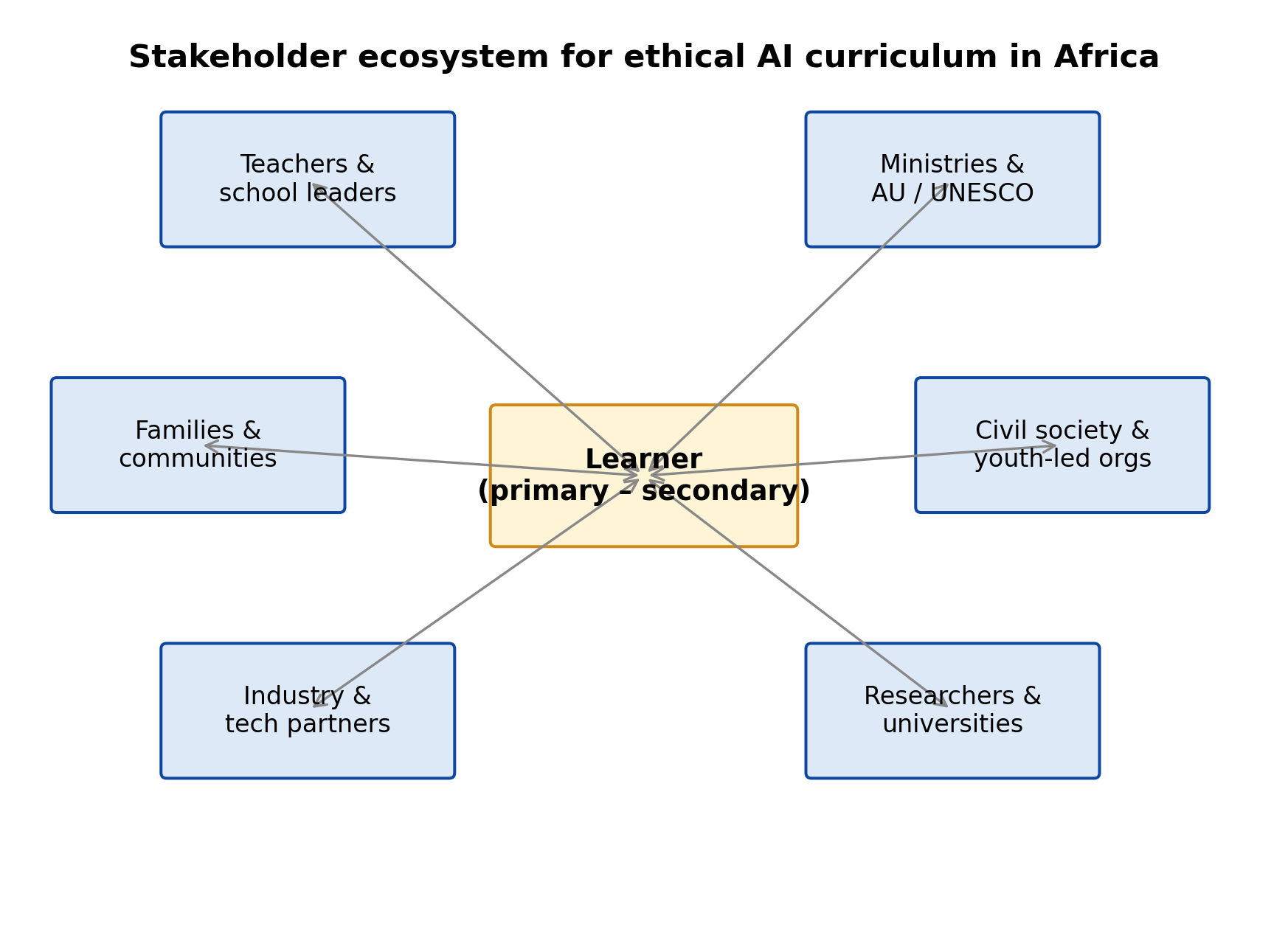}
\caption{Stakeholder ecosystem for ethical AI curriculum in Africa.}
\label{fig:ecosystem}
\end{figure}

\subsection{Comparative Policy Landscape}
\begin{table}[ht]
\centering
\caption{Illustrative comparative AI-in-education policy landscape across selected African countries.}
\label{tab:policy}
\scriptsize
\begin{tabularx}{\textwidth}{@{}lXXXX@{}}
\toprule
\textbf{Country} & \textbf{National AI strategy} & \textbf{Ethics framing} & \textbf{Education integration} & \textbf{Notes} \\ \midrule
Kenya & National AI Strategy 2025--2030~\citep{kenya2025ai,team4tech2025kenya} & Yes (explicit) & Embedded via CBC; DigiKen hubs & $\sim$35\% schools with internet; teacher PD limited \\
South Africa & Coding \& Robotics curriculum (selected grades) & Partial & Curriculum pilots & Strong research base \\
Rwanda & Smart Africa-aligned; Rwanda AI Policy & Yes & Targeted pilots & Centralised coordination \\
Ghana & AI Strategy (in development)~\citep{smartafrica2022ai} & Yes (emerging) & Planned & Active AI research community \\
Nigeria & National Digital Economy Policy \& Strategy~\citep{nitda2020ndeps} & Partial & Sector-wide digital literacy & Large-scale; uneven capacity \\
Cameroon & ICT in secondary education~\citep{cameroon2023ict} & Implicit & ICT integration & Francophone/anglophone context \\
Senegal & SN Num\'erique 2025 & Emerging & Sector-wide & Active civil society \\
Egypt & National AI Strategy & Yes & University focus & Arabophone context \\ \bottomrule
\end{tabularx}
\end{table}

\subsection{Existing African AI Education and Research Initiatives}
\begin{table}[ht]
\centering
\caption{Illustrative African AI-education and research initiatives relevant to framework implementation.}
\label{tab:initiatives}
\small
\begin{tabularx}{\textwidth}{@{}lXX@{}}
\toprule
\textbf{Initiative} & \textbf{Focus} & \textbf{Relevance to framework} \\ \midrule
Masakhane & NLP for African languages~\citep{nekoto2020masakhane,africanlp2024} & Translation, content adaptation, teacher training \\
Deep Learning Indaba & Pan-African ML research and capacity building & Mentorship, community of practice, curriculum content \\
Data Science Africa & Data science capacity across Africa & Teacher PD, datasets, case studies \\
Zindi & African data-science competitions & Real-world problems for capstone projects \\
UNESCO RAM (Southern Africa) & AI Readiness Methodology assessments~\citep{unesco2024ram} & National baselines; policy input \\
Smart Africa Alliance & Pan-African digital strategy~\citep{smartafrica2022ai} & Regional standards; scaling \\
African Observatory on Responsible AI & Research and policy on responsible AI & Ethics guidance; stakeholder convening \\
DigiKen (Kenya) & Digital innovation hubs~\citep{kenya2025ai,team4tech2025kenya} & Curriculum pilot sites; teacher training \\ \bottomrule
\end{tabularx}
\end{table}

\section{Planned Empirical Validation Programme}
\label{sec:empirical}
The framework in this paper is a hypothesis for practice. Translating it into evidence requires a mixed-methods empirical programme that is African-led, multi-country, and ethically governed.

\subsection{Research Design}
We propose a three-phase mixed-methods design combining a policy--practitioner Delphi study, a continental teacher survey, and classroom piloting in a small set of participating schools.

\subsection{Phase 1 --- Delphi Study}
A two-round modified Delphi study~\citep{hsu2007delphi} will convene 30--40 participants representing AI researchers, ethics scholars, ministry representatives, teacher educators, civil-society organisations, and youth-led digital-rights groups.

\subsection{Phase 2 --- Teacher Survey}
A structured teacher survey will be conducted across anglophone, francophone, lusophone, and arabophone African contexts, targeting $\ge$600 primary and secondary teachers across at least six countries.

\subsection{Phase 3 --- Classroom Piloting}
12--20 participating schools across 3--4 countries will pilot age-band-appropriate modules for one academic term.

\subsection{Ethical and Governance Considerations}
Research ethics approvals will be obtained from each participating institution and relevant national education authorities. Consent processes include age-appropriate assent for learners, informed consent from parents or guardians, and community-level agreement with school leadership. Data governance follows the UNESCO Recommendation on AI ethics~\citep{unesco2021recom} and national data-protection regulations. An African-led ethics advisory board supervises the programme.

\subsection{Roadmap and Timeline}
\begin{figure}[ht]
\centering
\includegraphics[width=0.95\textwidth]{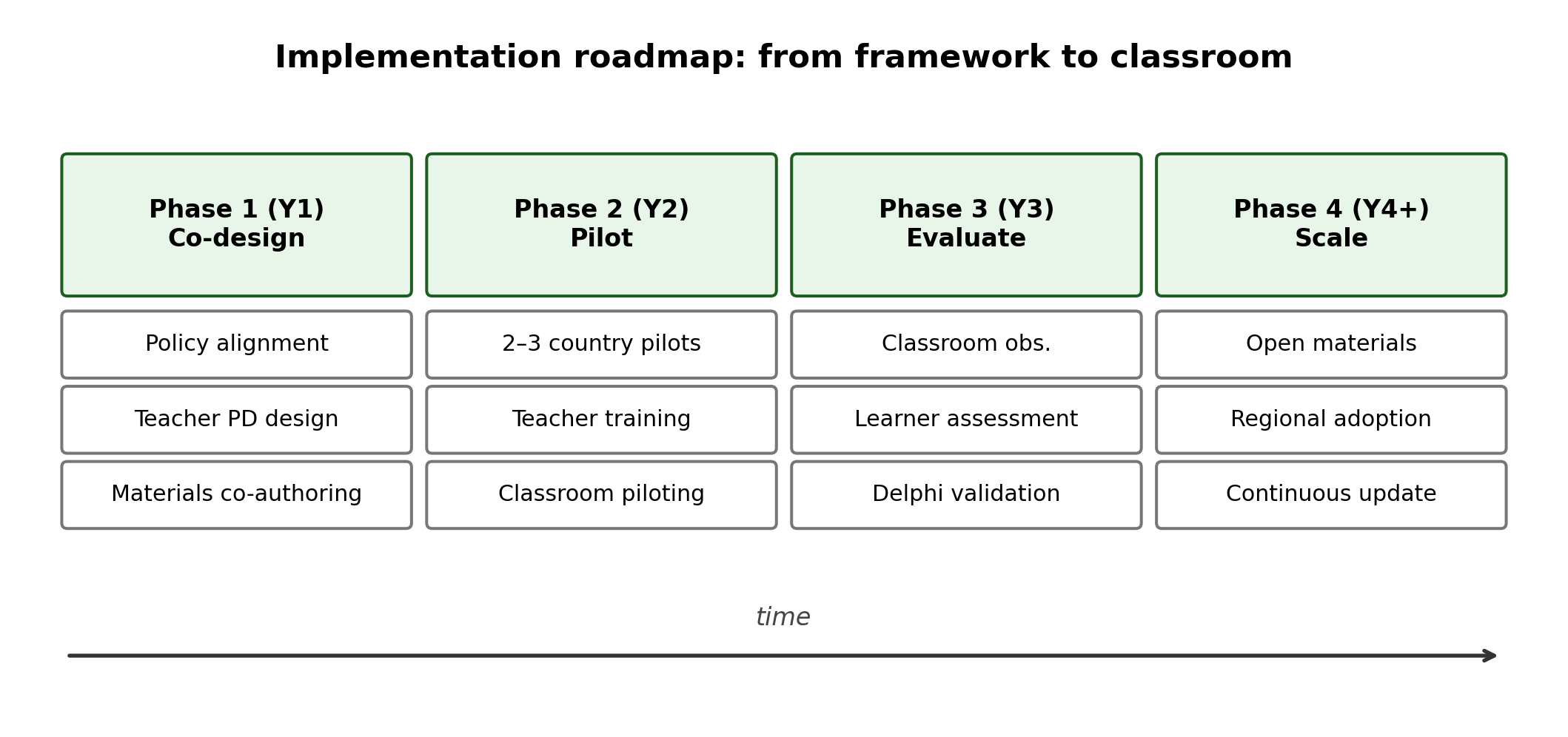}
\caption{Phased implementation roadmap from co-design through pilot, evaluation, and scale.}
\label{fig:roadmap}
\end{figure}

\section{Discussion}

\subsection{Barriers}
Four barriers recur: infrastructural disparities~\citep{kenya2025ai,williams2020equity}; thin teacher capacity~\citep{ouma2024kenya,williams2024teacher}; fragmented policy environments~\citep{au2024contai,smartafrica2022ai}; and under-resourced language and cultural localisation~\citep{nekoto2020masakhane,africanlp2024}.

\subsection{Opportunities}
Africa's youthful demographic is an asset, as is the vitality of African scholarship on AI ethics and decolonial thought. Regional integration initiatives offer platforms for shared curriculum resources. African AI research communities~\citep{nekoto2020masakhane,africanlp2024} already produce locally relevant content.

\subsection{Tensions}
We retain two productive tensions: between universal AI-ethics principles and African ethical traditions (Table~\ref{tab:ubuntu}); and between rapid uptake driven by workforce-readiness narratives and the slower, deliberative pedagogy that ethical reasoning requires.

\subsection{Barriers--Enablers Map}
\begin{figure}[ht]
\centering
\includegraphics[width=0.85\textwidth]{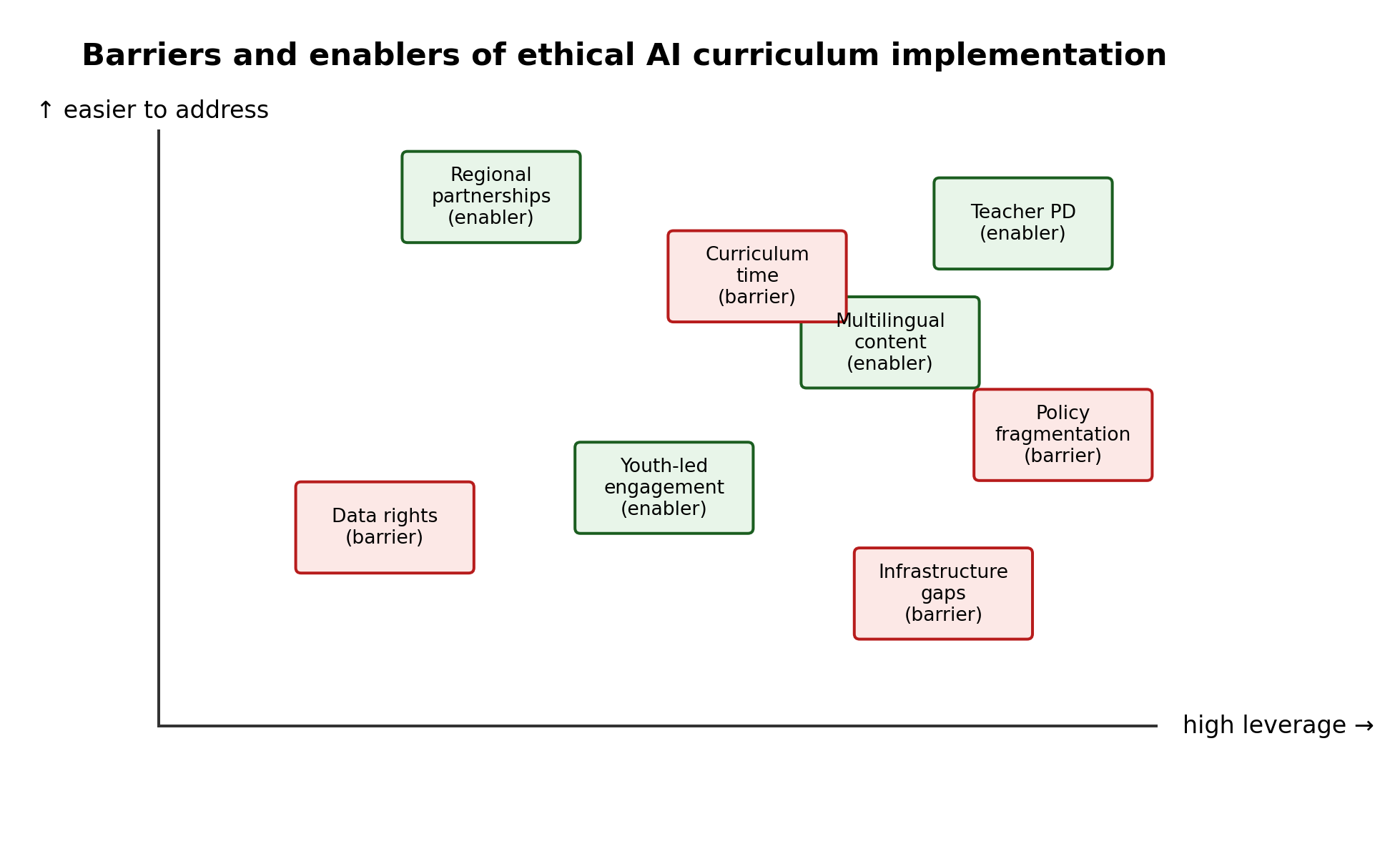}
\caption{Indicative barriers--enablers map. Items plotted to the right are higher-leverage; items plotted higher are easier to address.}
\label{fig:barriers}
\end{figure}

\subsection{Limitations of This Study}
As a conceptual synthesis, its claims are framework-level hypotheses for practice. Deployment must be preceded by country- and community-level co-design. The source base skews towards English-language scholarship; francophone, lusophone, and arabophone sources are under-represented.

\section{Recommendations}

\subsection{For Policymakers}
\begin{itemize}[leftmargin=*,itemsep=2pt]
\item Articulate an ethical AI education workstream within national AI and education strategies, aligned with the AU Continental AI Strategy~\citep{au2024contai} and UNESCO Recommendation~\citep{unesco2021recom}.
\item Align national standards with continental guidance while preserving space for national and linguistic adaptation.
\item Mandate data-protection and algorithmic-accountability provisions for any AI tool deployed in schools.
\item Invest in teacher professional development specifically on AI ethics.
\end{itemize}

\subsection{For Educators and School Leaders}
\begin{itemize}[leftmargin=*,itemsep=2pt]
\item Begin with low-stakes integration using Table~\ref{tab:bands}.
\item Prioritise unplugged and inquiry-based pedagogy where infrastructure is constrained.
\item Build communities of practice across schools.
\item Protect formative and reflective assessment for ethical competencies.
\end{itemize}

\subsection{For Civil Society and Youth-Led Organisations}
\begin{itemize}[leftmargin=*,itemsep=2pt]
\item Co-design curriculum materials with schools and communities.
\item Document and share local case studies of AI impact.
\item Advocate for transparent procurement, data-rights, and accountability standards.
\end{itemize}

\subsection{For International Partners}
\begin{itemize}[leftmargin=*,itemsep=2pt]
\item Fund African-led curriculum development and teacher training rather than exporting Global-North content.
\item Invest in translation, localisation, and open-licensed resources in African languages~\citep{nekoto2020masakhane,africanlp2024}.
\item Support empirical piloting through multi-year, multi-country partnerships with transparent governance.
\end{itemize}

\section{Conclusion and Future Research}
An African ethical AI curriculum cannot be assembled by translating Global-North content into local languages. It must be built on ethical principles interpreted through culturally responsive and decolonial pedagogy, grounded in African ethical traditions such as Ubuntu, and accountable to African communities. The framework offers a structured foundation: five AI-ethics principles made explicit through an Ubuntu-informed mapping, six guiding principles, four curriculum domains, an age-banded progression, five competency areas, and a coherent set of pedagogical strategies.

Future research will pursue the Delphi, survey, and classroom-pilot phases in Section~\ref{sec:empirical}, develop and openly licence curriculum materials in African languages in partnership with African NLP communities, study teacher professional-development models that scale ethical AI education without diluting its critical and culturally responsive core, and support youth-led research on how African learners themselves experience AI systems.

\section*{Declarations}
\textbf{Funding.} No external funding was received for this conceptual synthesis.

\textbf{Competing interests.} The authors declare no competing interests.

\textbf{Data availability.} This study is a conceptual synthesis. All policy documents and peer-reviewed sources cited are publicly available. Materials developed in the planned empirical programme will be released under open licences.

\textbf{Ethical approval.} Not applicable for this conceptual synthesis. Ethical approvals for the planned empirical programme will be obtained from relevant bodies in each participating country.

\textbf{Author contributions.} A.K.~Adedeji conceptualised the study and led the policy and pedagogical synthesis. F.~Tchakount\'{e} contributed to the analysis of francophone African policy contexts and to the theoretical framework. S.O.~Yusuff contributed to the analysis of AI ethics literature and to the implementation and recommendations sections. All authors drafted, reviewed, and approved the final manuscript.

\textbf{Acknowledgements.} The authors thank colleagues in African AI research and teacher-education communities whose published and informal conversations shaped this work.

\bibliographystyle{plainnat}
\bibliography{aicurr}

\end{document}